\documentclass[showpacs,amssymb,floatfix,twocolumn]{revtex4}
\usepackage{graphicx}
\usepackage{dcolumn}
\usepackage{graphics}
\usepackage{epstopdf}
\usepackage{pdfpages}
\usepackage{bm}
\begin{document}
\title{Robust bound states in the continuum in Kerr microcavity
embedded in photonic crystal
waveguide}
\author{Evgeny N. Bulgakov and Almas F. Sadreev}
\address{Institute of Physics, Academy of Sciences, 660036 Krasnoyarsk,
Russia}
\date{\today}
\begin{abstract}
We present a two-dimensional photonic crystal design with a microcavity of
four defect dielectric rods with eigenfrequencies residing
in the propagating band of directional waveguide. In the linear case for
tuning of material parameters of defect rods the nonrobust bound state in the continuum (BSC)
might occur. The BSC is a result of
full destructive interference of resonant monopole and quadrupole modes with the same parity.
A robust BSC arises in a self-adaptive way
without necessity to tune the parameters of the microcavity with the Kerr effect. Lack of the superposition
principle in nonlinear systems gives rise to coupling of the BSC with
injecting light. That forms a peculiar shape of isolated transmittance resonance
around BSC frequency.
We show if injecting light is switched off the BSC storages light that opens a way for
light accumulation.
\end{abstract}

\maketitle
\section{Introduction}
In 1929, von Neumann and Wigner \cite{neumann} predicted the
existence of discrete solutions of the single-particle
Schr\"odinger equation embedded in the continuum of positive
energy states, bound states in the continuum (BSC). Their analysis examined by Stillinger and Herrick
\cite{stillinger} long time was regarded as mathematical curiosity
because of certain spatially oscillating central symmetric
potentials. That situation cardinally changed when Friedrich and Wintgen \cite{friedrich}
in framework of two-level Fano-Anderson model formulated the
BSC  as a resonant state whose width tends
to zero as at least one physical parameter varies continuously
(see, also \cite{shahbazyan,volya,guevara,PRB}). Localization of
of the resonant states of open system, i.e., the  BSC
can be interpreted as destructive interference of two resonance states which occurs
for crossing of eigenlevels of the closed system \cite{PRB}. That accompanied by avoiding crossing
of the resonant states one of which transforms into the trapped state with vanishing width
while the second resonant state acquires the maximal resonance width (superradiant state
\cite{volya,PRB}).

The BSC phenomenon is a manifestation of wave interference similar
to the Aharonov-Bohm effect or the Anderson localization and is generic in all wave systems.
In particular Shipman and Venakides \cite{Shipman} predicted a symmetry protected
trapping of electromagnetic waves in
periodical array of dielectric rods. Two theoretical groups independently  presented
examples of the  BSC in photonics \cite{Shabanov,photonic}.
In Refs. \cite{Shabanov,Ndangali} the infinite periodic double dielectric gratings
 and two arrays of dielectric cylinders were considered where the BSC is localized in
direction cross to the arrays. In Ref. \cite{photonic} the photonic crystal (PhC) waveguide
with directional continuum in two-dimensional PhC of dielectric rods with
two off-channel optical microresonators was considered to show various types of the BSC.
In both systems
the BSC is the result of the Fabry-Perot mechanism for the BSC \cite{Kim,FanPRB,RS}
which is accompanied by the Fano resonance collapse in transmittance.
In forthcoming papers such photonic BSCs were experimentally observed
\cite{Lepetit,Segev,Longhi,Wei}. A realization of the BSC in the one-dimensional PhC
by an advanced digital grading
method was described in Ref. \cite{Prod}.
The BSC at surface of half infinite bulk system lays also a new concept for surface
states \cite{molina,Longhi,Wei1,Kivshar,Gallo}.

In this letter we present a PhC design of in-channel optical microcavity embedded into
the waveguide and show that it capable to realize the BSC as the result of destructive interference
of two resonant modes with the same parity decaying into the waveguide continuum.
However in the linear open systems the BSC occurs at the unique singular point of space of physical
parameters \cite{volya,ring,PRB} that
constitutes a difficulties for experimental visualization of the net BSC in PhC system.
First, it is necessary to vary material parameters of the microcavity in order to
approach to the BSC point. Second, the BSC is decoupled from the waveguide continuum
\cite{ring,PRB}  and therefore the BSC can not be probed by incoming  waves.

Our aim is to show that these difficulties can be overcome if to explore the Kerr
effect of the optical microcavity.
The BSC appears by self-adaptive way due to the Kerr shift of the dielectric constant
of the microcavity \cite{FAM} that transforms the BSC point into the BSC line
in the space of frequency and dielectric constant. Also  nonlinearity
lifts the principle of linear superposition to give rise to that injecting wave interacts
with the BSC. Therefore incoming light excites the BSC and forms novel type of complicated
response crucially different from Breit-Wigner or Fano type of resonances.
Thus the nonlinearity opens a new page in the BSC phenomenon \cite{molina,FAM} and
promises  novel nonlinear effects \cite{FAM,Shipman1,Ndangali1,degeneracy} in the
light transmission.
\section{Linear case}
\begin{figure}[ht]
\includegraphics[height=5cm,width=5cm,clip=]{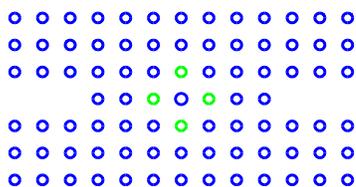}
\caption{PhC consists of a square lattice of GaAs rods
with linear refractive index $n_0=3.4$ and nonlinear refractive index
$n_2=1.5\times 10^{-13}cm^2/W$ at $\lambda=1.55m\mu$ and radius 0.18a in air shown by blue
open circles where. $a$ is the lattice unit. Four linear defect rods of the
same radius with dielectric constant  $\epsilon$  are shown by green bold circles.} \label{design}
\end{figure}

\begin{figure}[ht]
\includegraphics[height=8cm,width=8cm,clip=]{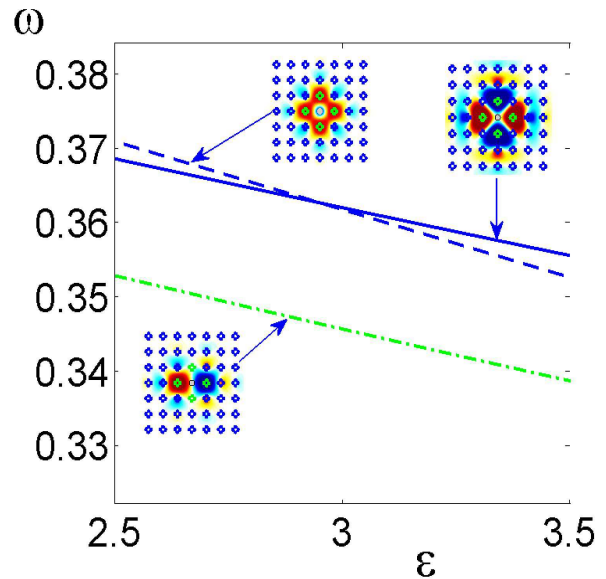}
\caption{(Color online) (a) Eigen-frequencies in unit of $2\pi c/a$ vs dielectric constant of the four
defect rods shown by open green circles in Fig. \ref{fig1}. Insets to them show profiles of
the eigenmodes.} \label{fig2}
\end{figure}

The layout of photonic crystal (PhC) system is shown in Fig. \ref{design} with
parameters given in Figure caption. The single row  of the rods is removed
from the PhC that forms a directional photonic waveguide which
supports a single band of guided TM mode spanning from the bottom
band edge 0.315 to the upper one 0.41 in terms of $2\pi c/a$
\cite{busch}. The TM mode has the electric field component
parallel to the infinitely long rods. Four linear defect rods of the
same radius with dielectric constant $\epsilon$ shown by green
open circles are placed at vertexes of square. The fifth central defect rod
made from the same GaAs material as the ghost rods of PhC is placed in the center of square.
These five defect rods form optical microcavity embedded into the PhC waveguide.
From both sides of the microcavity additional couple of rods are inserted in the waveguide
in order to diminish the coupling constant.
The eigenfrequencies of the cavity versus the dielectric constant $\epsilon$ are
plotted in Fig. \ref{fig2} which are accompanied by the eigenmodes \cite{busch}.
The  quadrupole-xy mode has  eigenfrequency beyond the
propagation band of PhC waveguide for considered range of interest of $\epsilon$ in
Fig. \ref{fig2}.

\begin{figure}[ht]
\includegraphics[height=5cm,width=6cm,clip=]{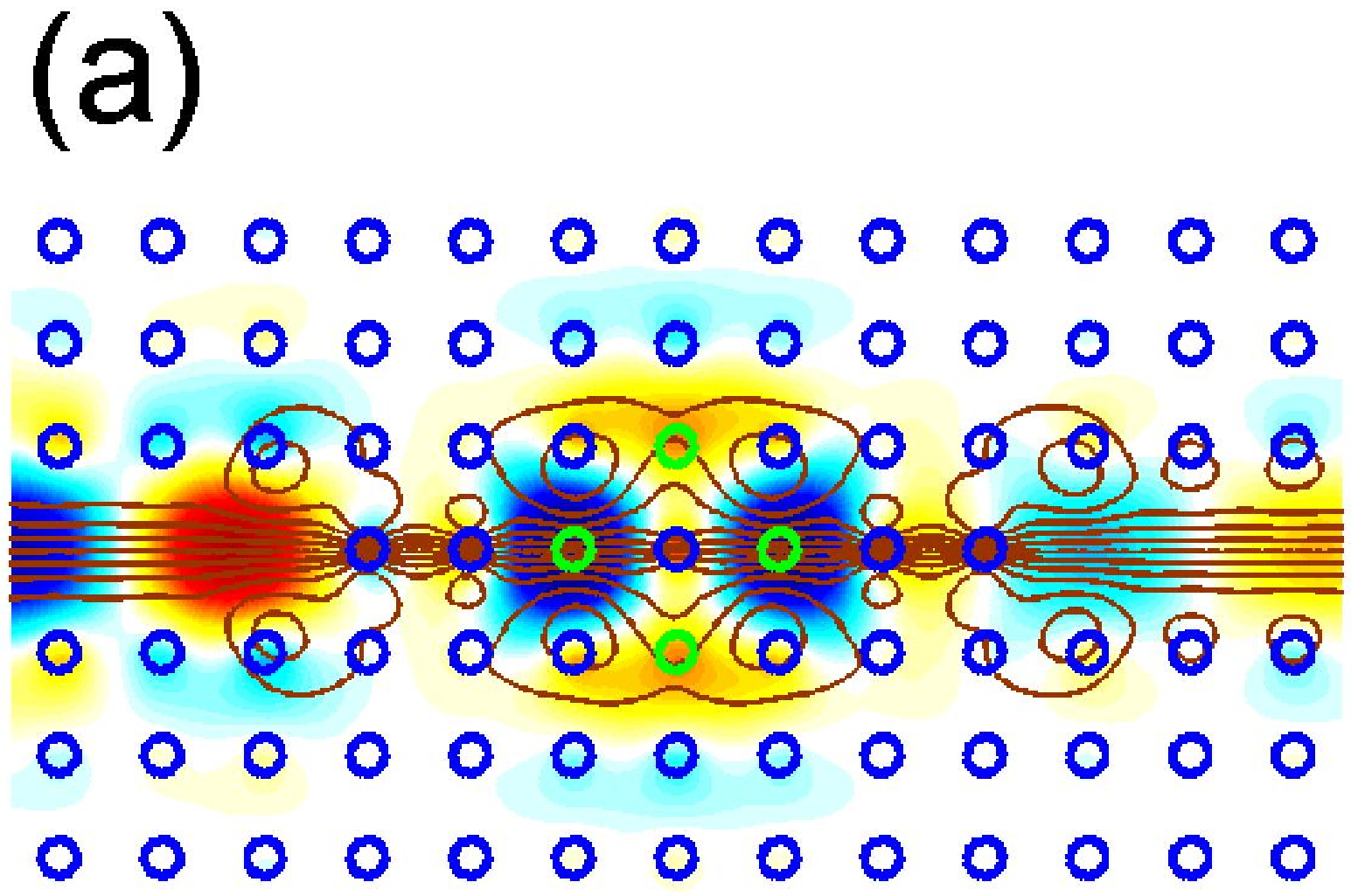}
\includegraphics[height=5cm,width=6cm,clip=]{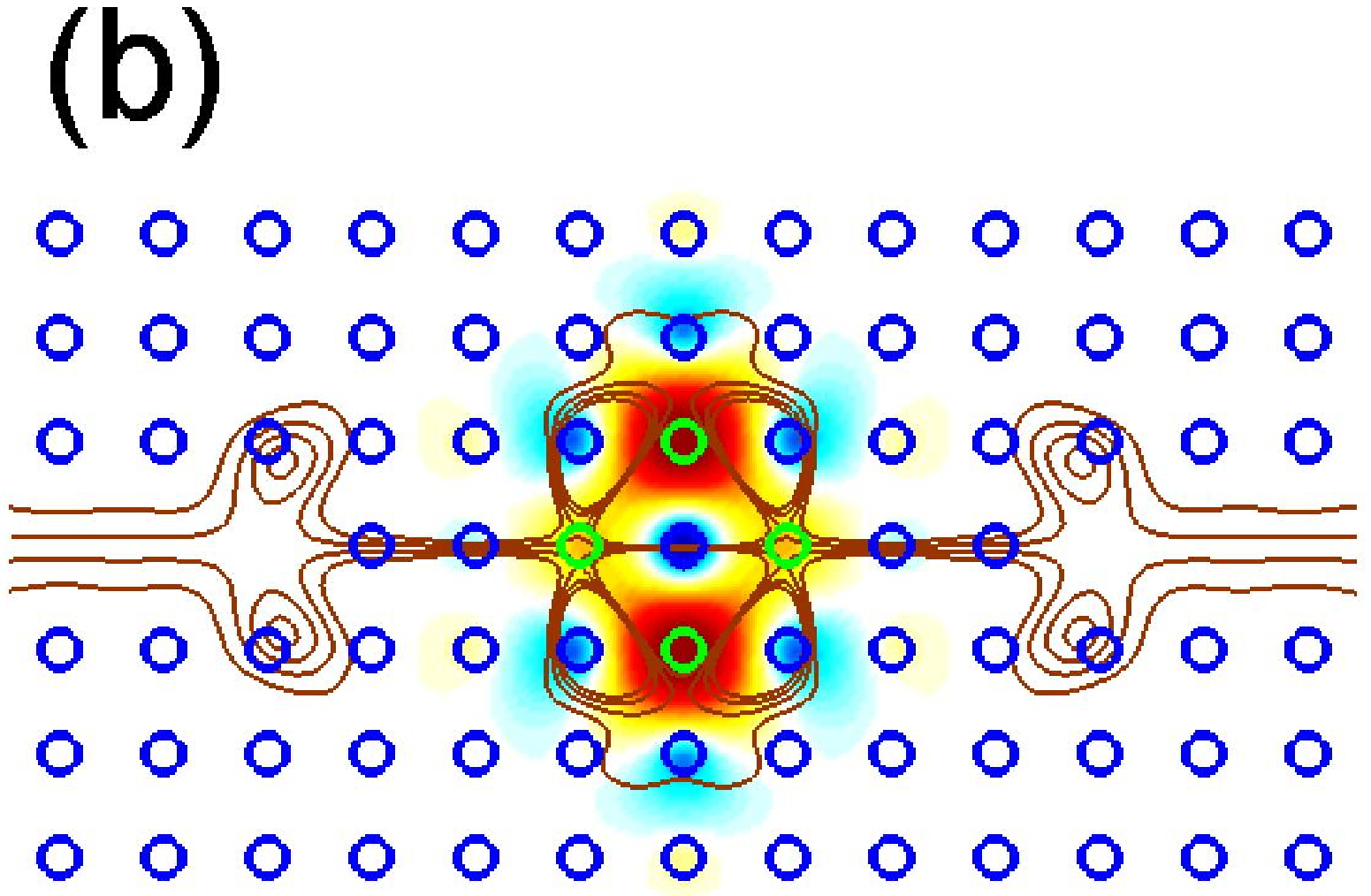}
\includegraphics[height=3.5cm,width=5cm,clip=]{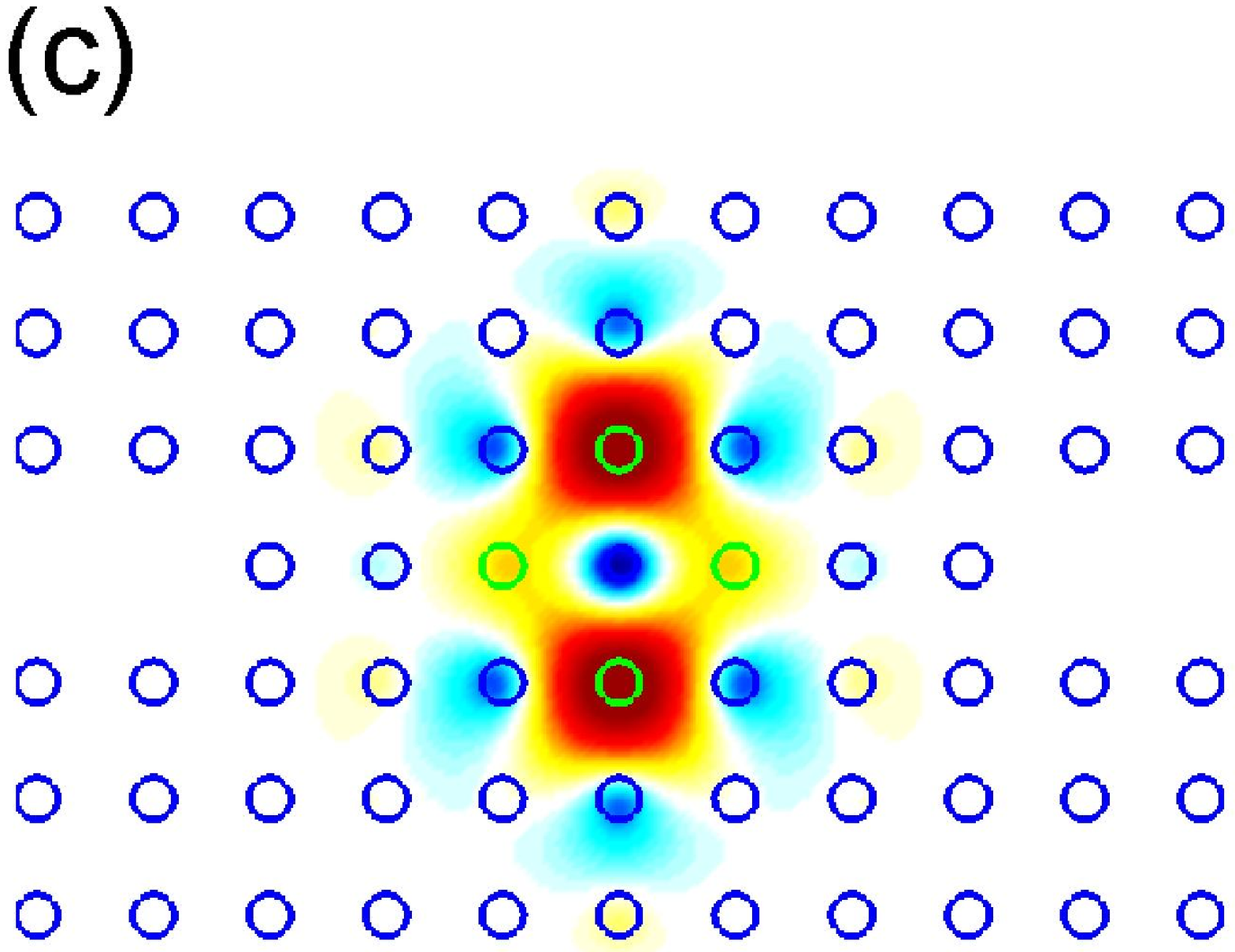}
\caption{The solutions  of Eq. (\ref{LS}) (real parts) for (a)
$\epsilon=3.01, \omega=0.3619$ and (b) $\epsilon=3.01, \omega=0.361755$.
Thin solid line shows current flows.
(c) BSC which is the eigenfunction of the effective Hamiltonian in Eq. (\ref{LS}) when
its complex eigenvalue becomes real (zero).} \label{fig1}
\end{figure}

\begin{figure}[ht]
\includegraphics[height=4.5cm,width=5cm,clip=]{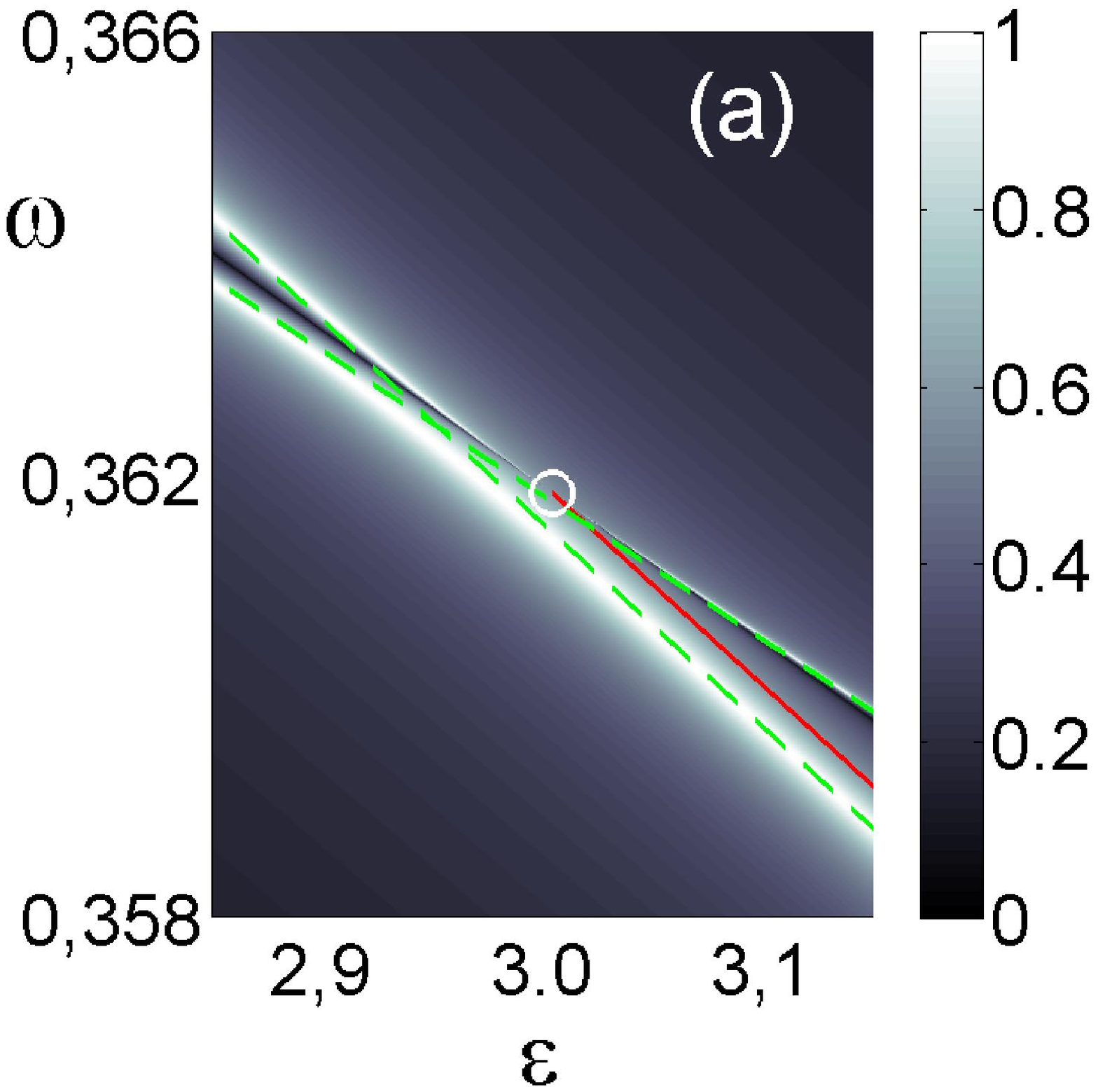}
\includegraphics[height=4cm,width=5cm,clip=]{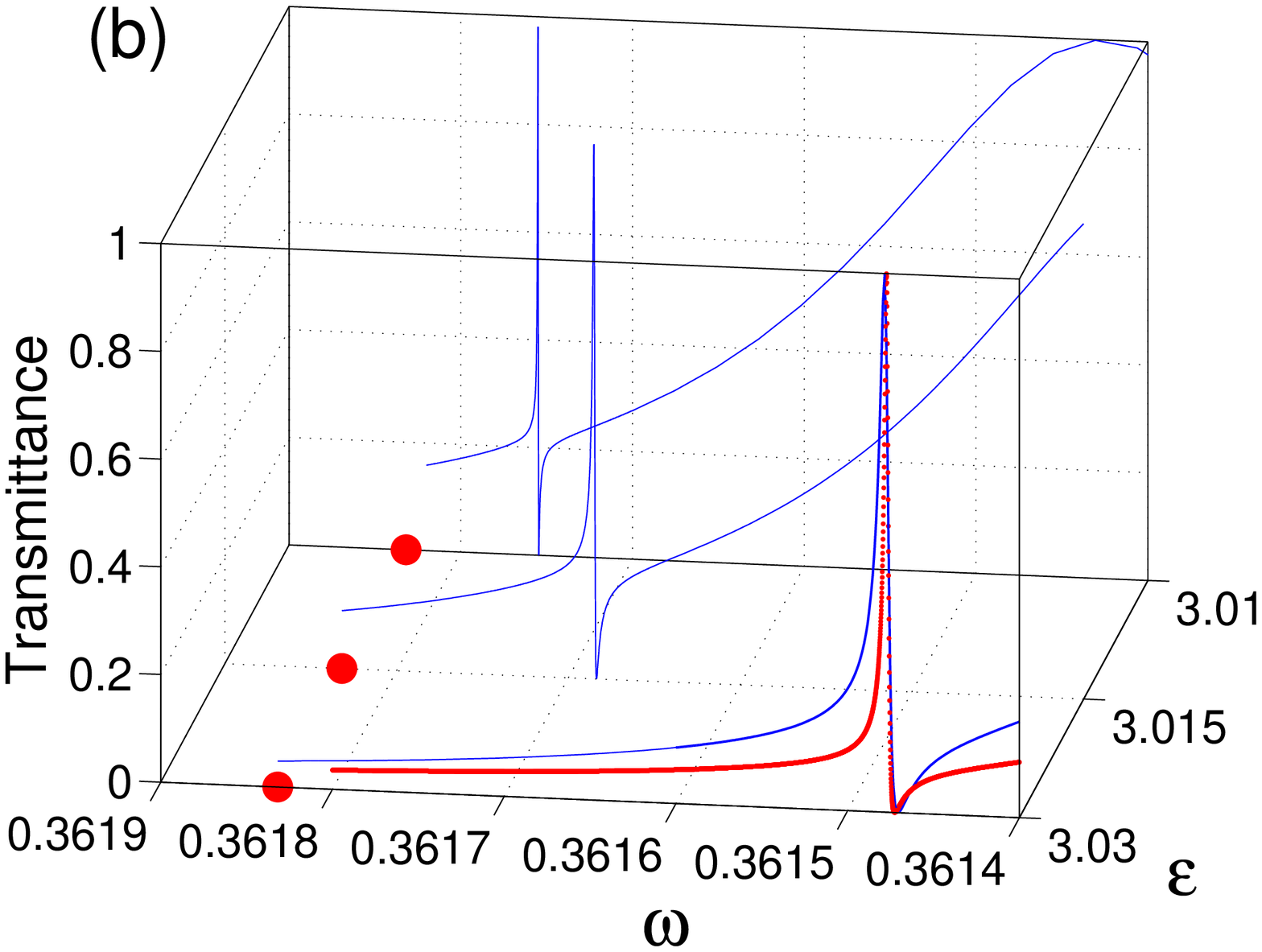}
\caption{(Color online) (a) Transmittance vs frequency of injected light and
dielectric constant of defect rods in the two-level approximation. The BSC point $\epsilon=3.004559, \omega=0.36183186$ marked
by white open circle. It is shown as this BSC point transforms into the line (red) if
to take into account the Kerr effect of central defect rod.  (b) transmittance vs
frequency in the vicinity of the BSC point. The three slice
correspond to $\epsilon=3.03, 3.015, 3.01$ respectively. Red closed circles mark the BSC frequency.
The results of computation for transmittance in PhC  are presented
by blue  line and CMT model by red.}
\label{fig3}
\end{figure}

The numerical procedure of solution of the Maxwell equations
is based on the Lippmann-Schwinger equation
\begin{equation}\label{LS}
[\hat{H}_{eff}(\omega)-\omega^2]\psi_S=\hat{\Gamma}\psi_{in},
\end{equation}
where $\hat{H}_{eff}(\omega)$ is the non-hermitian effective Hamiltonian which is resulted
by projection of the total space of the PhC system onto the inner space of the microcavity.
Respectively, the scattering function $\psi_S$
is  electric field directed along the rods in the microcavity, while the right-hand
expression in Eq. (\ref{LS}) shows as injected light amplitude
$\psi_{in}$ excites the microcavity through the coupling matrix $\hat{\Gamma}$.
We refer to Refs. \cite{photonic,T} for details in application to
the PhC. The complex eigenvalues of $H_{eff}$ have simple physical
meaning \cite{Ingrid}. Its real parts define resonant frequencies of the cavity,
and imaginary parts are responsible for resonance widths.
The unique BSC point can be hardly achieved experimentally.
Therefore it is important to show by which way to limit to the BSC point in order to reveal
the scattering state maximally close to the BSC.
In Fig. \ref{fig1} (a) and (b) we show solutions of Eq. (\ref{LS})  for two frequencies
and the dielectric constant of defect rods $\epsilon=3.01$.
The first frequency corresponds non resonant transmittance marked by star in Fig. \ref{fig3}
(b) with the corresponding solution shown in Fig. \ref{fig1} (a).
The second choice of frequency corresponds to the resonant transmittance $|t|=1$ marked by
rhombus in Fig. \ref{fig3} (b) with the corresponding solution shown in Fig. \ref{fig1} (b).
In Fig. \ref{fig1} (c)
the BSC $\psi_{BSC}$ is shown which is
eigenfunction of the non-hermitian effective Hamiltonian $H_{eff}(\omega_{BSC})\psi_{BSC}=
\omega_{BSC}^2\psi_{BSC}$ with real eigenfrequency $\omega_{BSC}$ of the BSC.
\cite{photonic,ring,PRB}. As seen from Fig. \ref{fig1} (b)
for approaching to the BSC point along the line $|t|=1$ reveals the BSC
provided that the parameter $\epsilon$ is close to the BSC point $\epsilon=3.004559$.
However
the difference between the  scattering wave function in Fig. \ref{fig1} (b)
and localized BSC function in Fig. \ref{fig1} (c) is that the BSC does not support current flows.
Fig. \ref{fig3} (b) shows  as the Fano resonance is collapsing for approaching to
the BSC magnitude of the dielectric constant.

Numerically, the dimension of the inner space of microcavity takes around of
thousands of sites per elementary cell in the finite difference scheme.
This numerical
routine can be enormously shortened if to use numerically calculated eigenmodes
and restrict ourselves by contribution of only two relevant (monopole and quadrupole-diag)
eigenmodes. That decimation procedure corresponds to the coupled mode theory (CMT) \cite{manolatou}
if to disregard radiation shifts in the effective Hamiltonian.
The stationary CMT equations have the following form
\begin{equation}\label{CMT}
    [\hat{H}^{(2)}_{eff}(\omega)-\omega]\left(\begin{array}{c}A_1\cr A_2\end{array}\right)=-i
    \left(\begin{array}{c}\sqrt{\gamma_1}\cr \sqrt{\gamma_2}\end{array}\right)\psi_{in}
\end{equation}
where
\begin{equation}\label{H2eff}
\hat{H}^{(2)}_{eff}(\omega)=\left(\begin{array}{cc} \omega_1-i\gamma_1 &
-u-i\sqrt{\gamma_1\gamma_2}\cr
-u-i\sqrt{\gamma_1\gamma_2} &\omega_2-i\gamma_2\end{array}\right),
\end{equation}
$\psi_{in}$ is the amplitude injected light, the subscripts 1, 2 refer to the
monopole and quadrupole-diag eigenmodes. We approximate their
eigenfrequencies as follows
$\omega_{1,2}=\omega_0\pm\Delta$ where the parameters were borrowed from numerics to be equal
$\Delta=0.0025367(\epsilon-2.9518), \omega_0=0.362443-0.01567683(\epsilon-2.9518)$.
Similar expansions take place if to vary the radius of defect rods. The
resonant widths $\gamma_1, \gamma_2$ were evaluated from
transmittance resonances  provided that
the resonances are not overlapped. As a result we obtained
$\gamma_1=3\cdot 10^{-5}, \gamma_2=1.3\cdot 10^{-4}$. The coupling constant between
the modes $u$ was evaluated by fitting of the BSC point in the CMT approach
to the BSC point evaluated in numerical solution of the Maxwell equations based on
Eq. (\ref{LS}) to obtain $u=1.768583\cdot 10^{-4}$.
The amplitude of transmittance is given by expression \cite{manolatou}
\begin{equation}
\label{trans}
t=\psi_{in}+\sqrt{\gamma_1}A_1+\sqrt{\gamma_2}A_1
\end{equation}

Comparison of the CMT approximation
with numerical solution of full equation (\ref{LS}) demonstrates good agreement.
For CMT approach the BSC point
can be found analytically from equation $Det[\hat{H}^{(2)}_{eff})-\omega]=0$
 which equals \cite{volya,PRB}
\begin{equation}\label{BSC}
    \omega_2-\omega_1=\frac{u(\gamma_2-\gamma_1)}{\sqrt{\gamma_1\gamma_2}},
    \omega_{BSC}=\omega_2+u\sqrt{\frac{\gamma_2}{\gamma_1}}.
\end{equation}
Among many intriguing properties of the BSC it's point is singular in parametric space
of $\omega$ and $\epsilon$ \cite{PRB,ring,Ndangali}. The transmittance and scattering wave
function crucially depend on a way in the space $\omega, \epsilon$ to limit to the BSC point in the vicinity of radius about the
coupling strengths. A feature of light transmittance shown in Fig. \ref{fig3}
is that line of zero transmittance touches  of unit transmittance at the BSC point
shown in Fig. \ref{fig3} by white open circle \cite{volya,PRB,Lepetit}.

\section{Nonlinear case}
Above consideration for the linear case shows that revelation of the BSC demands  fine
tuning of material parameter (dielectric constant or diameter) of defect rods in order to satisfy equation
for the BSC point (\ref{BSC}).
Therefore a probing of BSC features in light transmittance by injecting light of monochromatic laser
is a challenge for experiment.
We show that account of the Kerr effect can lift this problem making the
BSC point self adaptive without a tuning of material parameters of the defect rods \cite{FAM,Valle}.

In the vicinity of the BSC point light intensity is sufficiently large only in the microcavity.
Therefore it is enough to modify the effective Hamiltonian
$\hat{H}^{(2)}_{eff}\rightarrow \hat{H}^{(2)}_{eff}+\hat{V}$. The matrix elements
of perturbation $\hat{V}$ in the two-level approximation equal  \cite{SB}
\begin{equation}\label{Vmn}
V_{mn}=-\frac{(\omega_m+\omega_n)}{4N_m}\int
d^2\vec{r}\delta\epsilon(\vec{r})E_m(\vec{r}) E_n(\vec{r}), m, n=1,2,
\end{equation}
where $E_m(\vec{r})$ are the eigenmodes of the linear microcavity shown in Fig. \ref{fig2}
with normalization
\cite{soljacic}
\begin{equation}\label{nor}
    N_m=\int d^2\vec{r}
    \epsilon_{PhC}E_m^2(\vec{r})=\frac{a^2}{cn_2},
\end{equation}
$\epsilon_{PhC}$ is the dielectric constant of whole defectless PhC.
\begin{equation}\label{deltaeps}
\delta\epsilon(\vec{r})=\frac{n_0cn_2|E(\vec{r})|^2}{4\pi}\approx
\frac{n_0cn_2|\sum_mA_mE_m(\vec{r})|^2}{4\pi}
\end{equation}
is the nonlinear contribution to the dielectric constant of the defect rod with
instantaneous Kerr nonlinearity,
After substitution of matrix (\ref{Vmn}) into the CMT equation (\ref{CMT})
we obtain the nonlinear system of quations for the mode amplitudes $A_m$.

\begin{figure}[ht]
\includegraphics[height=4cm,width=4cm,clip=]{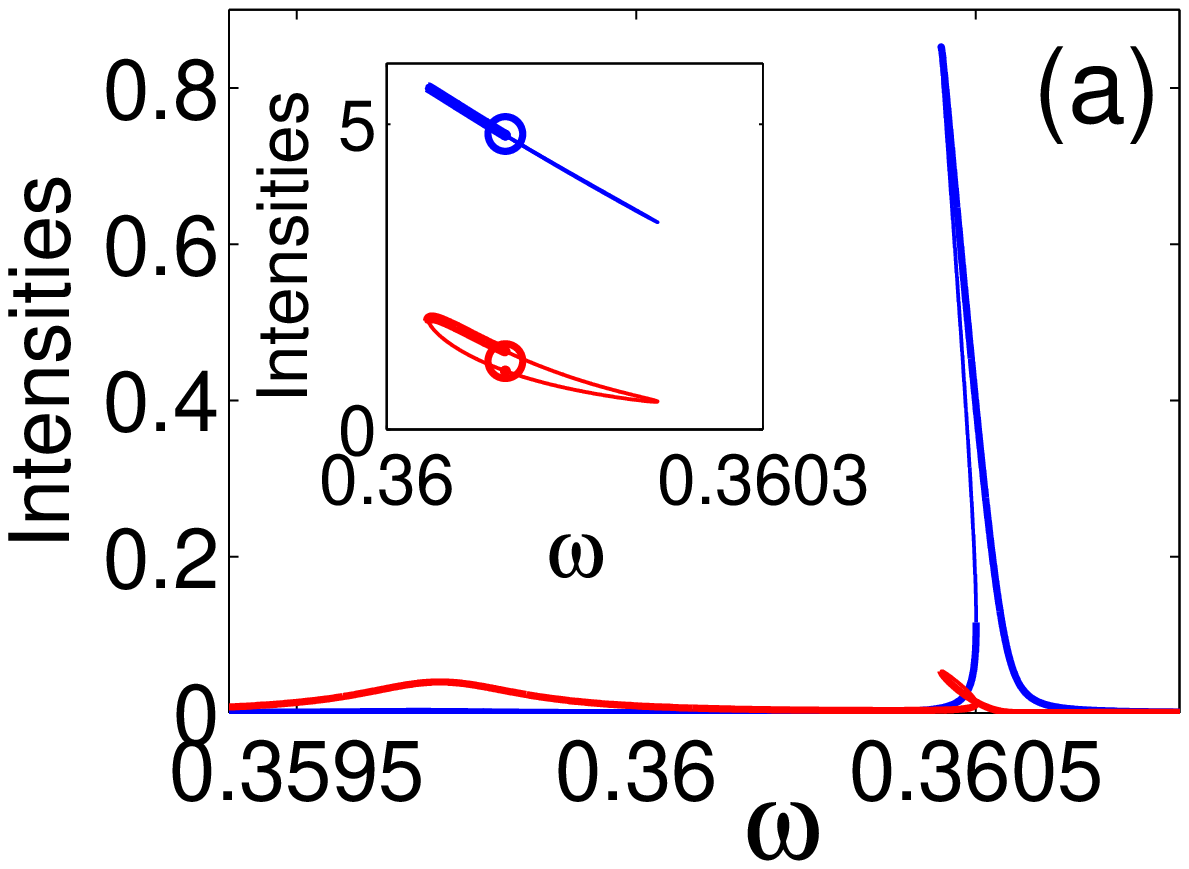}
\includegraphics[height=4cm,width=4cm,clip=]{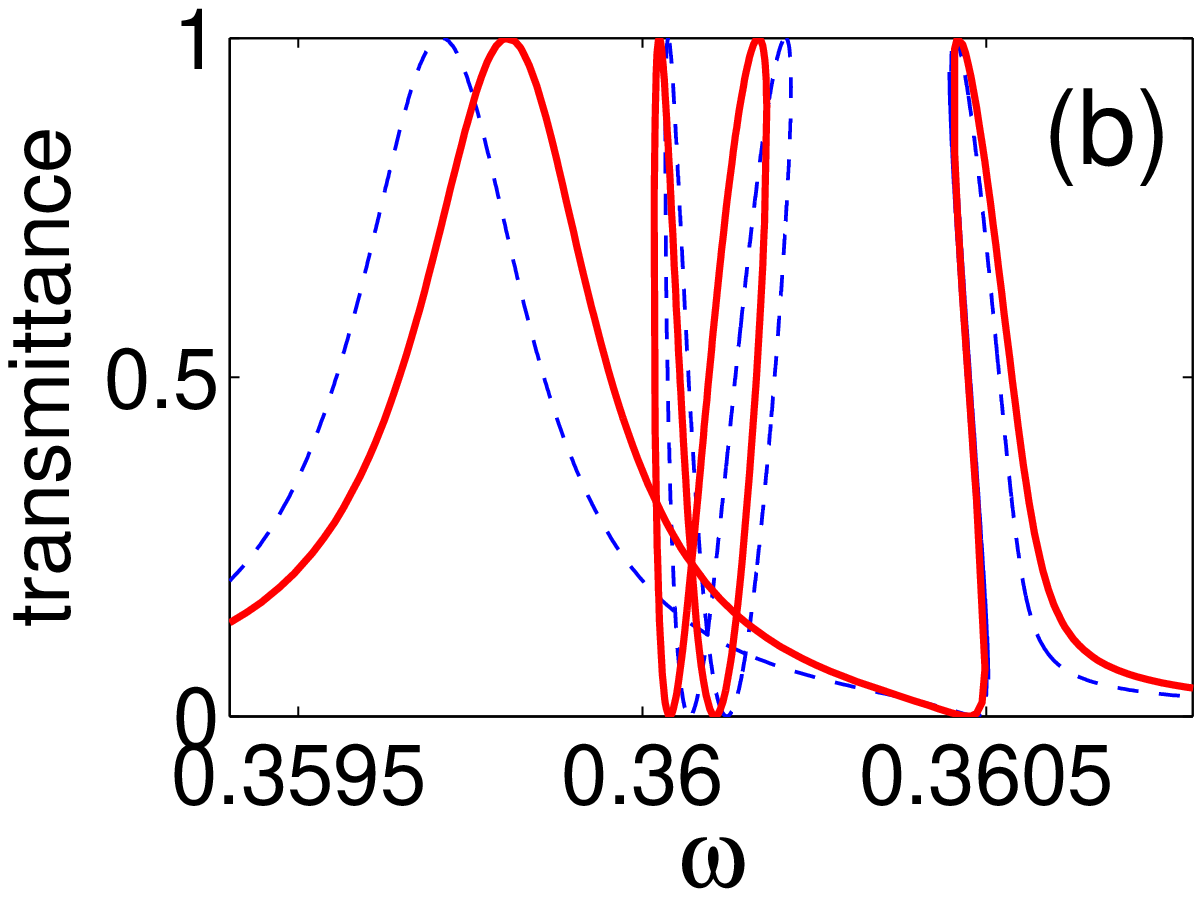}
\caption{(Color online) (a) Intensities of excitation of the monopole mode $|A_1|^2$
(blue lines)
and quadrupole mode $|A_2|^2$ (red lines). The basic window show the first family which
inherits the linear case and inset shows the BSC family.
Thin lines response for unstable solutions and thicker lines show stable solutions.
(b) Transmittance vs frequency of injected light with the amplitude $\psi_{in}=0.0025$
with account of Kerr effect with
$\lambda_{11}=0.0001$. The dielectric constant of vertex rods $\epsilon=3.1$.
Transmittance calculated from the nonlinear CMT equations (\ref{CMT})
is shown by blue dash line and transmittance calculated from  the nonlinear Maxwell equations is shown by red solid line.}
\label{fig4}
\end{figure}

Two factors substantially weakens the nonlinear contribution into the
quadrupole-diag mode. First, as seen from Fig. \ref{fig2} two nodal lines of the quadrupole mode
go through the central defect rod. Second, $\gamma_1\ll \gamma_2$ to give rise to inequality
$\lambda_{11}I_1\ll \lambda_{22}I_2$
where \cite{T}
\begin{equation}\label{lam11}
\lambda_{mn}=\frac{c^2n_2^2}{a^2}\int E_m^2(x,y)E_n^2(x,y)d^2\vec{r}.
\end{equation}
Therefore we can restrict ourselves by the nonlinear shift of the
first monopole mode frequency $\omega_1\rightarrow\omega_1+V_{11}=\omega_1-\lambda_{11}|A_1|^2$ only.
Then Eq. (\ref{BSC}) gives that the BSC point is achieved if
the intensity of the monopole excitement equals
\begin{equation}\label{BSCint}
\lambda_{11}I_{1c}=\lambda_{11}|A_1|^2=\omega_2-\omega_1-
\frac{u(\gamma_2-\gamma_1)}{\sqrt{\gamma_1\gamma_2}},
\end{equation}
with the BSC frequency defined in Eq. (\ref{BSC}). From equation for the BSC
$Det[\hat{H}^{(2)}_{eff})]=0$ and second equation in CMT equations (\ref{CMT})
we have an equality $(\omega_{BSC}-\omega_2+i\gamma_2)A_2+(u+i\sqrt{\gamma_1\gamma_2})A_1=0$
which defines the intensity of the quadrupole-diag mode at the BSC point
\begin{equation}\label{I2c}
    I_{2c}=|A_{2c}|^2=\frac{\gamma_1}{\gamma_2}I_{1c}.
\end{equation}
Both BSC intensities are marked in Fig. \ref{fig4} (a) by open circles.

After substitution of the nonlinear term $\omega_1\rightarrow\omega_1-\lambda_{11}|A_1|^2$
into Eq. (\ref{CMT}) and solving of self-consistent nonlinear equations we obtain two
different families of the solutions \cite{FAM}. The first family of solutions inherits the linear case
and for small injecting power has typical resonance behavior for  the mode intensities
$|A_1|^2, |A_2|^2$ shown in Fig. \ref{fig4} (a) in basic window.
Those mode (monopole) which has smaller resonant width undergoes larger excitation and larger
typical decline to the left because of negative contribution of the nonlinear term to the first
monopole mode.
The second BSC family of solutions are loops centered at the BSC point
(\ref{BSCint}) shown in inset of Fig. \ref{fig4} (a). A stability of the solutions are notified by thicker lines.
When the amplitude of injecting light  $\psi_{in}$ tends to zero
the size of loops is shrinking
to the BSC points marked by open circles. The transmittance calculated by Eq. (\ref{trans})
is plotted in Fig. \ref{fig4} (b) for both
families and clearly reflects the frequency behavior in Fig. \ref{fig4} (a). The transmittance
on the basis of full range nonlinear Maxwell equations is
plotted in Fig. \ref{fig4} (b) by solid red line
to demonstrate good agreement.
The BSC solutions exist for a whole range of
linear diffractive index as plotted by red line in Fig. \ref{fig3} (a) that
indeed makes the BSC for nonlinear optical microcavity flexible relative to choice of material
parameter. By the terminology proposed in Ref. \cite{Shipman}
we call such a BSC as the robust BSC.

\begin{figure}[ht]
\includegraphics[height=5cm,width=5cm,clip=]{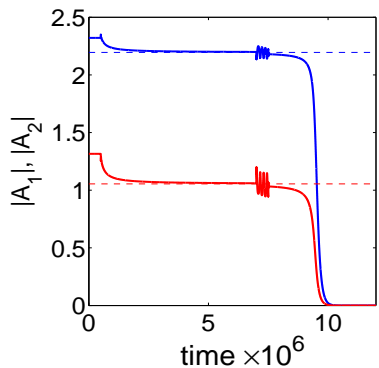}
\caption{(Color online) Time evolution of mode amplitudes after the injecting
light with amplitude power  $\psi_{in}=0.0025$ was switched off at time $0.5\times 10^6$
and then light impulse of duration $0.5\times10^6$ was applied  at the time $7\times 10^6$.
Dash lines show the BSC amplitudes
given by Eqs. (\ref{BSCint}) and (\ref{I2c}).}
\label{fig5}
\end{figure}
One can see from Fig. \ref{fig4} (a) that intensities of mode excitation at the BSC solution
substantially exceed the intensities for the solution inherited the linear
case, at least, for small injecting amplitudes.
The values $I_{1c}, I_{2c}$ around which the intensities are centered can be
enhanced by increasing of the linear refractive index of four defect rods or
by decreasing of the nonlinear refractive index of the central defect rod
as Eqs. (\ref{BSCint}) and (\ref{I2c}) show. That prompts to use the BSC solution for
storage of light. Indeed, Fig. \ref{fig5} shows time evolution of the mode amplitudes after
injecting light was switched off. The amplitudes fastly evolve from current values to the BSC values
given by Eq. (\ref{BSCint}) and shown by dash lines in Fig. \ref{fig5}.
Also it is easy to release this
accumulated energy by short impulse of light injected into the PhC waveguide.
These results open a way for light energy accumulation and release.

{\bf Acknowledgments}.
 The paper was partially supported by RFBR grant 13-02-00497 and grant of RSCF 14-12-00266.
 We acknowledge discussions with Chia Wei Hsu and D.N. Maksimov.

\end{document}